\documentclass[journal]{IEEEtran}
\usepackage{cite}
\usepackage[cmex10]{amsmath}
\usepackage{amssymb}
\usepackage{url}
\usepackage{graphicx} 
\begin{document}
\title{Synchronization-Free Delay Tomography\\Based on Compressed Sensing}

\author{
Kensuke~Nakanishi,
Shinsuke~Hara,
Takahiro~Matsuda,
Kenichi~Takizawa,
Fumie~Ono,
and~Ryu~Miura

\thanks{K. Nakanishi and S. Hara are with
Graduate School of Engineering, Osaka City University, Osaka, 5588585, Japan
email:\{nakanishi.k@c., hara@\}info.eng.osaka-cu.ac.jp}%

\thanks{T. Matsuda is with Graduate School of Engineering, Osaka
University, Osaka, 5650871, Japan email: matsuda@comm.eng.osaka-u.ac.jp
}%
\thanks{K. Takizawa, F. Ono, and R. Miura are with Wireless Network
Research Institute, National Institute of Information and Communications
Technology (NICT),
Kanagawa, 2390847, Japan
email:\{takizawa, fumie, ryu\}@nict.go.jp
}
\thanks{S. Hara and T. Matsuda are also with Wireless Network Research
Institute, NICT, Kanagawa, 2390847, Japan}
}


\maketitle

\begin{abstract}

Delay tomography has so far burdened source and receiver measurement
nodes in
a network with two requirements such as path establishment and clock
synchronization between them. In this letter, we focus on the clock
synchronization problem in delay tomography and propose a
{\em synchronization-free delay tomography} scheme. The proposed scheme
selects a
path between source and receiver measurement nodes as a reference
path,
which results in a loss of equation in a conventional delay tomography
problem. However, by utilizing compressed sensing, the proposed scheme
becomes robust to the loss. Simulation experiments confirm that
the proposed scheme works comparable to a conventional delay tomography
scheme in networks with no clock synchronization
between source and receiver measurement nodes.
\end{abstract}


\section{Introduction}
\IEEEPARstart{D}{elay}
tomography means to estimate internal link delays in a network by means
of measuring end-to-end path delays~\cite{Coates2002}.
When an active measurement procedure is used,
source measurement nodes transmit probe packets to receiver measurement
nodes,
and the end-to-end path delays are computed
from the differences between the transmission and reception times of the
probe packets.

So far, delay tomography has burdened source and receiver measurement
nodes with two requirements;
path establishment and clock synchronization between them.
In this letter,
we focus on the clock synchronization problem in delay tomography and
propose a {\em synchronization-free delay tomography} scheme.
Although there have been several works for the path establishment
problem~\cite{Takemoto2013},
to the best of the authors' knowledge, the clock synchronization problem
has not been studied so far.

The proposed scheme utilizes compressed sensing,
which is a promising technique because it can reduce the number of paths
between source and receiver measurement nodes~\cite{Firooz2010,Xu2010},
and identifies bottleneck links without any clock synchronization
mechanism between them.
In the proposed scheme,
we construct a {\em differential routing matrix}
by setting a path between source and receiver measurement
nodes as a reference path,
so
it results in a loss of equation for
the delay tomography problem.
Since compressed sensing is robust to this problem,
however, the proposed scheme works comparable to a conventional delay
tomography scheme in networks with no clock synchronization
between source and receiver measurement nodes.

The proposed scheme has a significant benefit in various network
environments
especially in wireless networks such as wireless sensor networks,
in which electronic components of nodes are sometimes too untrustable
to meet the requirement of clock synchronization in terms of accuracy
and complexity.
In addition, it gives an insight to general problems of compressed
sensing in which measurement process has an unknown bias.

\section{Preliminary for Compressed Sensing}

First, we define the $\ell_p$ norm ($p \geq 1$) of
a vector $\mathbf{x} = [x_1~x_2~\cdots~x_N]^\top \in \mathcal{R}^{N}$
as
\begin{equation}
\| \mathbf{x} \|_p = \Bigl( \sum_{i=1}^{N} |x_i|^p \Bigr)^{\frac{1}{p} },
\end{equation}
where $\top$ denotes the transpose operator.

Now, we assume that
through a matrix $\mathbf{A} \in
\mathcal{R}^{M \times N}$ ($M <N$),
a vector
$\mathbf{y} = [y_1~y_2~\cdots~y_M]^\top \in \mathcal{R}^{M}$ is obtained
for a vector $\mathbf{x}$
as $\mathbf{y}=\mathbf{A}\mathbf{x}$.
When utilizing
compressed sensing,
whether or not one can recover a sparse vector $\mathbf{x}$ from
$\mathbf{y}$
depends on the mathematical property of $\mathbf{A} $.
Here,
we define
the mutual coherence $\mu(\mathbf{A})$, which can
provide guarantees of the recovery of the sparse vector,
as
\begin{equation}
\label{eqn:mutualcoherence}
\mu(\mathbf{A}) = \max_{1 \leq j,j' \leq N,j \neq j'}
\frac
{ |\mathbf{a}_j^\top \mathbf{a}_{j'} | }
{ \| \mathbf{a}_j \|_2 \| \mathbf{a}_{j'} \|_2},
\end{equation}
where $\mathbf{a}_j$ and $\mathbf{a}_{j'}$ are
the $j$-th and $j'$-th column vectors of
$\mathbf{A}$, respectively.
If
\begin{equation}
k < \frac{1}{2} \Bigl( 1 + \frac{1}{\mu(\mathbf{A})} \Bigr),
\label{k}
\end{equation}
then there exists at most one vector $\mathbf{x}$ which has at most {\it k}
nonzero components~\cite{Elad2010}.

\section{Conventional Delay Tomography}
\label{systemmodel}
Let $\mathcal{G} = (\mathcal{V}, \mathcal{E})$ denote an
undirected network, where $\mathcal{V}$ and $\mathcal{E}
\subset \mathcal{V} \times \mathcal{V}$ denote sets of nodes and links,
respectively.
In addition, let $s \in \mathcal{V}$
and $r \in \mathcal{V}$
denote source and receiver measurement nodes,
respectively.

We assume that there are only two measurement nodes, which is natural in
practical environments
because it is difficult to deploy many measurement nodes, especially in
large-scale networks.
Therefore,
we define $\mathcal{W} = \{\mathit{path}_{s,r}^{(l)};~l = 1, 2, \ldots,
|\mathcal{W}|\}$ as a subset of all paths
from $s$ to $r$,
where $\mathit{path}_{s,r}^{(l)} = \{(s, v_{s,r}^{(l,1)}),
(v_{s,r}^{(l,1)}, v_{s,r}^{(l,2)}), \ldots,
(v_{s,r}^{(l,|\mathit{path}_{s,r}^{(l)}|-1)}, r)\} \subset
\mathcal{E}$
represents the $l$-th path in $\mathcal{W}$ and
$v_{s,r}^{(l,m)}\in \mathcal{V}\setminus \{s,r\}$
$(m=1,\ldots,|\mathit{path}_{s,r}^{(l)}|-1)$
are intermediate nodes in the path.
Furthermore, we reformulate $\mathcal{W}$ and $\mathcal{E}$ as
$\mathcal{W} =
\{w_1, w_2, \ldots, w_I\}$ and $\mathcal{E} = \{e_1,
e_2, \ldots, e_J\}$, respectively, where $I = |\mathcal{W}|$ and $J =
|\mathcal{E}|$ denote the numbers of paths and links, respectively,
and define $d_{e_j}$ as the delay over $e_j$ ($j = 1,2, \ldots, J$).
Finally, we define a binary matrix $\mathbf{A} \in \{0,1\}^{I\times J}$
as the {\em routing matrix}
of $\mathcal{W}$
(each row of the matrix is a path), i.e.,
its ($i$,$j$) components are set to $a_{ij} = 1$ if $e_j \in w_i$,
and $a_{ij} = 0$ otherwise.

Conventional delay tomography has been discussed on ideal networks
which have no clock synchronization error
between source and receiver measurement nodes.
In this case,
a packet transmitted from $s$ on a path $w_i$ ($i = 1,
2, \ldots, I$) is successfully received at $r$ with total
delay $D_{w_i}= \sum_{e_j \in w_i} d_{e_j}$, so
defining {\em measurement vector} $\mathbf{y} =
[y_1~y_2~\cdots~y_I]^\top$ and {\em link delay vector}
$\mathbf{x} = [x_1~x_2~\cdots~x_J]^\top$~as
\begin{equation}
y_i =D_{w_i}= \sum_{e_j \in w_i} d_{e_j},
~~x_j =d_{e_j},
\end{equation}
and by using $\mathbf{A}$, we naturally obtain
the following matrix/vector equation:
\begin{equation}
\label{eqn:system}
\mathbf{y} = \mathbf{A}\mathbf{x}.
\end{equation}
In this letter, we assume that link states are stationary, i.e., link
delays do not change while the proposed scheme is applied.

When we are interested in identification of a limited number of
bottleneck links with larger delays, we can apply compressed sensing.
Namely,
by attributing the delays only to the bottleneck links,
we can approximate the elements of $\mathbf{x}$ corresponding to
smaller link delays to be zero,
so we can assume that $\mathbf{x}$ is approximately a sparse vector.

To calculate the mutual coherence of $\mathbf{A}$,
by picking up the {\it j}-th and {\it j'}-th column vectors from
$\mathbf{A}$,
we define the partial matrix as
\begin{equation}
\mathbf{A} _{jj'} =
\left[
\begin{array}{cc}
\mathbf{a}_j & \mathbf{a}_{j'}
\end{array}
\right].
\label{original}
\end{equation}
When we obtain $\tilde{\mathbf{A}} _{jj'}$ by
swapping any two row vectors of $\mathbf{A}$,
from (\ref{eqn:mutualcoherence}),
we can see $\mu(\tilde{\mathbf{A}}_{jj'})=\mu(\mathbf{A}_{jj'})$.
So by repeating the swap, $\mathbf{A} _{jj'} $ leads to
\begin{equation}
\widetilde{\mathbf{A}} _{jj'} =
\left[
\begin{array}{cc}
\mathbf{d}_{jj'} & \overline{\mathbf{d}}_{jj'} \\
\mathbf{s}_{jj'} & \mathbf{s}_{jj'}
\end{array}
\right]
,
\label{swap}
\end{equation}
where
$\mathbf{d}_{jj'}$ and $\mathbf{s}_{jj'}$ are
row vectors with adequate dimensions, respectively,
and $\overline{(\cdot)}$ denotes the bit-reverse operator.
This means that,
by changing the order of the elements of
$\mathbf{a}_j$ and $\mathbf{a}_{j'}$,
they can be rearranged into two column vectors which
have different elements in the upper part whereas
the same elements in the lower part.

If $\frac{1}{3} \leq \mu(\mathbf{A}) < 1$, namely, $k = 1$,
$\mathbf{A}$ is referred to as {\em $1$-identifiable} matrix.
In the following, we assume
$\mathbf{A}$ is $1$-identifiable.
In this case, from (\ref{swap}),
for $1 \leq j,j' \leq J,~j \neq j'$
\begin{equation}
{\rm dim} (\mathbf{d}_{jj'}) \ne 0
\label{dim}
\end{equation}
is held.
In brief, $\mu(\mathbf{A}) < 1$ means that
any two absolute column vectors of $\mathbf{A}$ are not the same
vectors.

\section{Proposed Differential Delay Tomography}

When the clock of the receiver measurement node is not synchronized to
that of the
source measurement node,
the measured total delay is contaminated with a synchronization error.
That is,
defining the clock synchronization error as $\Delta$,
the real measurement vector $\mathbf{z}=[z_1~z_2~\cdots~z_I]^\top$
should be written as
\begin{equation}
\mathbf{z} = \mathbf{y}+\Delta \cdot \mathbf{1},
\label{eq:prin1}
\end{equation}
where $\mathbf{1}$ and $\mathbf{y}$ are the all-one vector
and the true measurement vector, respectively.
Thus, we have
\begin{equation}
\mathbf{z} = \mathbf{A}\mathbf{x}. \label{eq:prin2}
\end{equation}
From (\ref{eq:prin1}) and (\ref{eq:prin2}),
we can see that
the conventional delay tomography does not work at all
unless $\Delta$ is estimated.

Now,
in order to get rid of the synchronization error completely,
we define the {\it r}-th component $z_r$ of $\mathbf{z}$ and
the {\it r}-th row of $\mathbf{A}$ as the {\em reference
component} and
the {\em reference row}, respectively.
By subtracting the reference component and the reference row from all
the other components and all the other rows, respectively, we have a new
{\em differential measurement vector} $\mathbf{z}^{(r)} \in
\mathbb{R}^{I-1}$ and
a new {\em differential routing matrix} $\mathbf{A}^{(r)} \in
\{-1,0,1\}^{(I-1) \times J}$ as
\begin{equation*}
\label{diffvec}
\mathbf{z}^{(r)}= [z_1-z_r~z_2-z_r~\cdots ~z_i-z_r~\cdots ~z_I-z_r]^\top,
\end{equation*}
\begin{equation*}
\mathbf{A}^{(r)} = \left[ \begin{array}{cccc}
a_{11}-a_{r1} & a_{12}-a_{r2} & \cdots & a_{1J}-a_{rJ}\\
a_{21}-a_{r1} & a_{22}-a_{r2} & \cdots & a_{2J}-a_{rJ} \\
\vdots & \vdots & \ddots & \vdots \\
a_{i1}-a_{r1} & a_{i2}-a_{r2} & \cdots & a_{iJ}-a_{rJ}\\
\vdots & \vdots & \ddots & \vdots \\
a_{I1}-a_{r1} & a_{I2}-a_{r2} & \cdots & a_{IJ}-a_{rJ}\\
\end{array} \right].
\end{equation*}
As a result, we have a new matrix/vector equation which does not
contain the synchronization error as
\begin{equation}
\mathbf{z}^{(r)}=\mathbf{A}^{(r)}\mathbf{x}.
\end{equation}
Note that the link delay vector $\mathbf{x}$
does not change.

In the proposed scheme,
the trade-off for the asynchronism is a loss of equation.
However, compressed sensing is a method to obtain a unique solution from
an underdetermined linear system,
so it is still simply applicable for the
proposed scheme.

To discuss recoverability of a sparse
vector $\mathbf{x}$ by means of the mutual coherence of $\mathbf{A}^{(r)}$,
let us remind $\mathbf{A}$ is assumed to be $1$-identifiable.
Due to the limitation of
the topology,
if $\mathbf{A}$ has a rearranged partial matrix
$\widetilde{\mathbf{A}}_{jj'}$ whose ${\rm dim} (\mathbf{d}_{jj'})$ equals $I$,
then $\mu (\mathbf{A}^{(r)})$ always equals $1$,
that is, $k$ of $\mathbf{A}^{(r)}$ always equals $0$.
Otherwise, $\mu(\mathbf{A}^{(r)})$ is still less than $1$, regardless of the
reference row,
that is, there is no loss of capacity in terms of the mutual coherence
property given by (\ref{k}).
\begin{IEEEproof}
See the Appendix.
\end{IEEEproof}

\begin{figure}[ttt]
\centering
\includegraphics[width=0.25\textwidth]{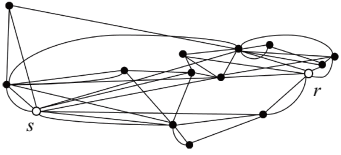}
\caption{Network Topology $1$ with 15 nodes and 44 links.}
\label{fig:topology}
\end{figure}

\begin{figure}[ttt]
\centering
\includegraphics[width=0.15\textwidth]{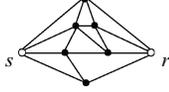}
\caption{Network Topology $2$ with 8 nodes and 16 links.}
\label{fig:topology2}
\end{figure}

\begin{table}[t]
\begin{center}
\caption{tested routing matrices}
\begin{tabular}{|c|c|c|c|}\hline
\raisebox{0.7ex}{Matrix} & \raisebox{0.7ex}{Size} &
\raisebox{0.7ex}{Topology} & \shortstack{Mutual \\ Coherence}\\ \hline
$\mathbf{P}$ & $20 \times 44$&Fig.$1$&$0.707$\\ \hline
$\mathbf{Q}$ & $25 \times 44$&Fig.$1$&$0.707$\\ \hline
$\mathbf{R}$ & $30 \times 44$&Fig.$1$&$0.707$\\ \hline
$\mathbf{S}$ & $8 \times 16$ &Fig.$2$&$0.816$ \\ \hline
$\mathbf{T}$ & $10 \times 16$&Fig.$2$&$0.816$\\ \hline
$\mathbf{U}$ & $12 \times 16$&Fig.$2$&$0.667$\\ \hline
\end{tabular}
\label{original_matrices}
\end{center}
\end{table}

\begin{table}[t]
\begin{center}
\caption{differential matrices}
\begin{tabular}{|c|c|c|c|c|}\hline
\raisebox{0.7ex}{Matrix} & \raisebox{0.7ex}{Size} & \shortstack{Original
\\ Matrix} & \shortstack{Reference \\ Row($\ell_1$norm)} &
\shortstack{Mutual \\ Coherence} \\
\hline
$\mathbf{P}^{(2)}$ & $19 \times 44 $& $\mathbf{P }$&$2nd~(2) $&
$0.944$\\ \hline
$\mathbf{Q}^{(2)}$ & $24 \times 44 $& $\mathbf{Q }$&$2nd~(2) $& $0.884 $\\
\hline
$\mathbf{R}^{(2)}$ &$ 29 \times 44 $& $\mathbf{R}$ & $2nd~(2)$ & $0.841$\\
\hline
$\mathbf{S}^{(2)}$& $7 \times 16 $&$\mathbf{ S }$& $2nd~(2)$ & $0.845
$\\ \hline
$\mathbf{T}^{(2)}$& $9 \times 16 $& $\mathbf{T }$& $2nd~(2)$ & $0.875$\\
\hline
$\mathbf{U}^{(2)}$& $11 \times 16 $& $\mathbf{U}$ & $2nd~(2)$& $0.843$\\
\hline
$\mathbf{R}^{(3)}$ & $29 \times 44 $& $\mathbf{R}$ & $3rd~(3)$ &$ 0.906 $\\
\hline
$\mathbf{R}^{(4)} $ &$ 29 \times 44 $& $\mathbf{R }$& $4th~(4)$ &
$0.964 $\\ \hline
$\mathbf{R}^{(5)}$ & $29 \times 44$ & $\mathbf{R}$ & $5th~(5)$ &
$0.964$\\ \hline
$\mathbf{R}^{(11)}$& $29 \times 44$ & $\mathbf{R}$ & $11th~(11)$ & $0.983
$\\ \hline
$\mathbf{R}^{(14)} $& $29 \times 44$ & $\mathbf{R}$ & $14th~(14)$ &
$0.983$\\ \hline
\end{tabular}
\label{differential_matrices}
\end{center}
\end{table}

\section{Performance Evaluation}
\label{sec:simulation}
\subsection{Simulation Environment}

In this section, we evaluate the performance of the proposed
differential delay tomography scheme by C++ simulation experiments.
Fig.~\ref{fig:topology} and \ref{fig:topology2} show the network
topologies with $44$ links and $16$ links for the performance evaluation,
respectively.
In both the network topologies,
there are a source measurement node $s$ and a receiver measurement node
$r$ whose clocks are not synchronized.
We construct routing matrices in reference to
a method discussed for sparsity-constrained network tomography
in \cite{Takemoto2013},
and measure the delays between $s$ and $r$
using an active measurement procedure.

Link delay inference based on (\ref{eqn:system}) can be classified into
several models~\cite{Xia2006}.
Since we are interested only in
the capacity of differential routing matrices as compared to
their original routing matrices,
we assume that the link delay is considered unknown but constant.
In more detail,
{\it k} links are selected and assigned a
delay of $10$ ms to denote that they are congested,
whereas
all the other links in the network are assumed to have i.i.d.
exponentially
distributed delays with average $0.05$ ms to denote that these links do
not undergo congestion~\cite{Xia2006}.

As an implementation of compressed sensing,
we employ an $\ell_1$-$\ell_2$
optimization~\cite{Zibulevski2010,Matsuda2011},
and evaluate the {\em k-identifiability ratio} $R=N^{(k)} / {}_{J} C_k$,
where $k$ and $N^{(k)}$
denote
the number of congested links
and the number of the congested link sets which
can be
identified from the routing matrix and the measurement vector, respectively.

Finally,
Table~\ref{original_matrices} and \ref{differential_matrices} show
the tested routing matrices and the differential matrices, respectively.
Note that in Table~\ref{differential_matrices}, the $\ell_1$norm of the
reference row
vector corresponds to the number of links over the selected reference
and the reference row is arranged so as to make its order
identical to its $\ell_1$norm.

\begin{figure}[ttt]
\centering
\includegraphics[width = .70\linewidth ]{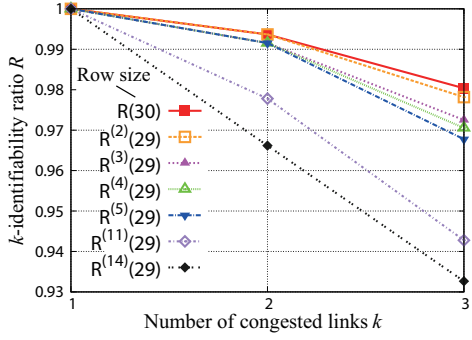}
\caption{{\it k}-identifiability ratio $R$ vs. the number of congested
links {\it k};
dependency on the number of links in the reference path
for network topology 1.}
\label{result_R}
\end{figure}

\begin{figure}[ttt]
\centering
\includegraphics[width = .70\linewidth ]{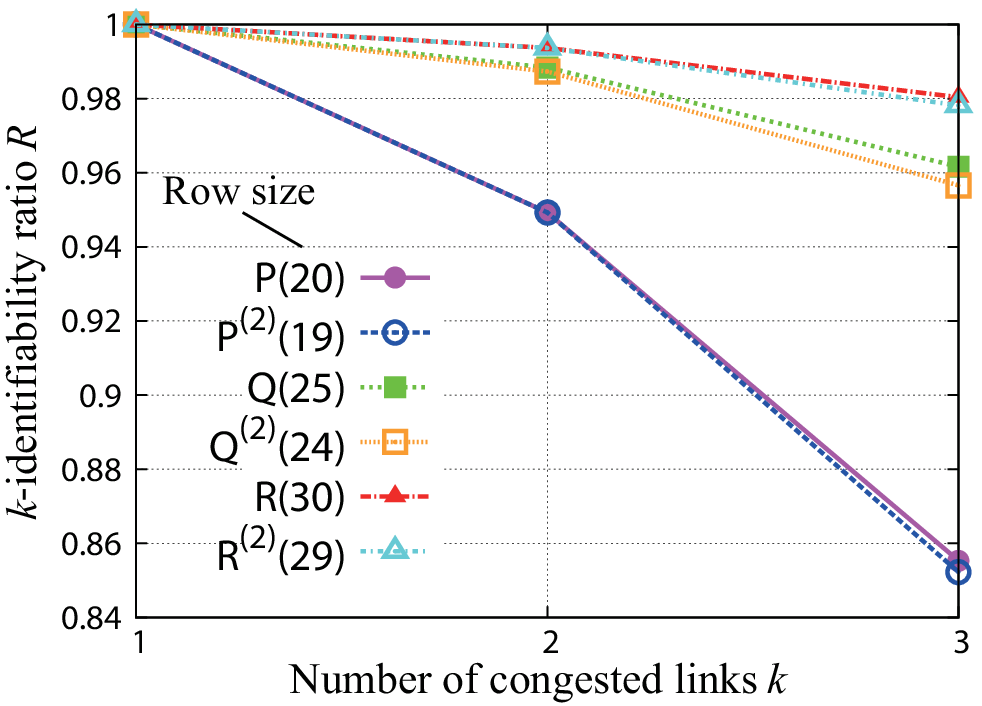}
\caption{{\it k}-identifiability ratio $R$ vs. the number of congested
links {\it k};
effect of the row dimension of routing matrix for network topology $1$. }
\label{result1}
\end{figure}

\begin{figure}[ttt]
\centering
\includegraphics[width = .70\linewidth ]{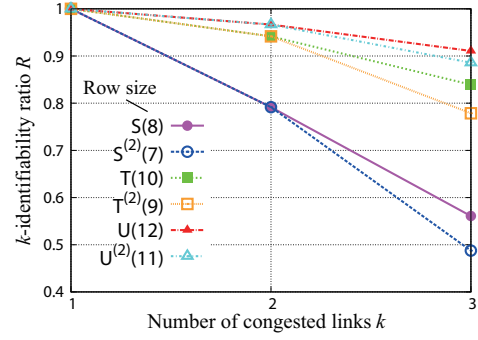}
\caption{{\it k}-identifiability ratio $R$ vs. the number of congested
links {\it k};
effect of the dimension of routing matrix for network topology $2$.}
\label{result2}
\end{figure}

\subsection{Simulation Results}
\label{sec:results}

Figs.~\ref{result_R}, \ref{result1}, and \ref{result2} show the
{\it k}-identifiability ratio $R$ vs. the number of congested links {\it k}.
To evaluate the capacity of the differential routing matrix,
it is meaningful to compare its performance with that for its original
routing matrix,
although the conventional delay tomography scheme does not work at all
in the
networks where the clock of $s$ is not synchronized to that of $r$.
Therefore, for comparison purpose,
the three figures contain the performance of the conventional delay
tomography scheme
when the clock of $s$ is assumed to be synchronized to that of $r$.
From Fig.~\ref{result_R}, we can see that the shorter path tends to be
more acceptable as the reference for the proposed scheme.
On the other hand, from Figs.~\ref{result1} and \ref{result2},
since the tested matrices have
different row sizes which correspond to the number of delay measurements,
we can see that
the more delay measurements are more advantageous in terms of performance
improvement for the proposed and original schemes.
For each of given numbers of delay measurements,
the proposed scheme,
in which
a path giving the highest performance is appropriately selected as the
reference,
performs comparable to the original scheme.

Consequently,
in the two networks,
because of the clock asynchronism between
source and receiver measurement nodes,
we have to select a path as the reference
in the propose scheme, which leads to a
loss of information.
However, the information of the reference can be deleted not completely
but partially.
If an appropriate reference is selected,
the differential delay
tomography scheme can effectively identify the congested links.

\section{Conclusion}
\label{sec:conclusion}
In this letter, we proposed a differential delay tomography scheme
which enables us to infer link delays without clock synchronization
between source and receiver measurement nodes
in a network.
We theoretically proved that the differential routing matrix preserves
the mathematical property of the 1-identifiability of its original routing
matrix
and evaluated the performance of the proposed scheme by the simulation
experiments.
Some technical issues remain in the proposed scheme.
Particularly, we have to propose an efficient reference selection method.
Since the issues are beyond the scope of this letter, we leave them as
future works.

\appendix
We define
the {\it r}-th row of $\mathbf{d}_{jj'}$ and $\mathbf{s}_{jj'}$
as $d^{(r)}_{jj'} \in \{0,1\}$ and $s^{(r)}_{jj'} \in \{0,1\}$,
respectively,
and the matrices obtained by deleting
the {\it r}-th row from
$\mathbf{d}_{jj'}$ and $\mathbf{s}_{jj'}$
as $\mathbf{d}^{(-r)}_{jj'}$ and $\mathbf{s}^{(-r)}_{jj'}$, respectively.
Incidentally $\mathbf{A}^{(r)}$ represents the differential matrix
of $\mathbf{A}$.\\
\noindent
1) If ${\rm dim} (\mathbf{d}_{jj'})=I$, that is,
$\widetilde{\mathbf{A}} _{jj'}=
[\mathbf{d}_{jj'}~~\overline{\mathbf{d}}_{jj'}]$:
\begin{eqnarray*}
\widetilde{\mathbf{A}}^{(r)} _{jj'}
& = &
[\mathbf{d}^{(-r)}_{jj'}-d^{(r)}_{jj'}\cdot \mathbf{1}
~~\overline{\mathbf{d}}^{(-r)}_{jj'}-\overline{d}^{(r)}_{jj'} \cdot
\mathbf{1}]
\nonumber \\
& = & \begin{cases}
~[ -\overline{\mathbf{d}}^{(-r)}_{jj'}~~\overline{\mathbf{d}}^{(-r)}_{jj'} ]
& (d^{(r)}_{jj'} = 1), \\
~[ \mathbf{d}^{(-r)} _{jj'}~~-\mathbf{d}^{(-r)} _{jj'} ] &
(d^{(r)}_{jj'} = 0).
\end{cases}
\end{eqnarray*}
Consequently, the two absolute column vectors of
$\widetilde{\mathbf{A}}^{(r)}_{jj'}$ have no different element,
so $\mu(\mathbf{A}^{(r)}) = 1$.
\\
2) Otherwise:\\
a) If the {\it r}-th row is selected from $\mathbf{d}_{jj'}$
then
\begin{eqnarray*}
\widetilde{\mathbf{A}}^{(r)} _{jj'}
& = &
\left[
\begin{array}{cc}
\mathbf{d}^{(-r)}_{jj'}-d^{(r)}_{jj'}\cdot \mathbf{1}
& \overline{\mathbf{d}}^{(-r)}_{jj'}-\overline{d}^{(r)}_{jj'} \cdot
\mathbf{1}\\
\mathbf{s}_{jj'}-d^{(r)}_{jj'}\cdot \mathbf{1}
& \mathbf{s}_{jj'}-\overline{d}^{(r)}_{jj'}\cdot \mathbf{1}
\end{array}
\right] \nonumber \\
& = &
\begin{cases}
~\left[
\begin{array}{cc}
-\overline{\mathbf{d}}^{(-r)}_{jj'}
& \overline{\mathbf{d}}^{(-r)}_{jj'} \\
-\overline{\mathbf{s}}_{jj'}
& \mathbf{s}_{jj'}
\end{array}
\right] & (d^{(r)}_{jj'} = 1),\\[15pt]
~\left[
\begin{array}{cc}
\mathbf{d}^{(-r)} _{jj'}
& -\mathbf{d}^{(-r)} _{jj'} \\
\mathbf{s}_{jj'}
& -\overline{\mathbf{s}} _{jj'}
\end{array}
\right] & (d^{(r)}_{jj'} = 0).
\end{cases}
\end{eqnarray*}
b) If the {\it r}-th row is selected from $\mathbf{s}_{jj'}$
then
\begin{eqnarray*}
\widetilde{\mathbf{A}}^{(r)}_{jj'}
& = &
\left[
\begin{array}{cc}
\mathbf{d}_{jj'}-s^{(r)}_{jj'}\cdot \mathbf{1}
& \overline{\mathbf{d}}_{jj'}- s^{(r)}_{jj'}\cdot \mathbf{1}\\
\mathbf{s}^{(-r)}_{jj'}-s^{(r)}_{jj'}\cdot \mathbf{1}
& \mathbf{s}^{(-r)}_{jj'}-s^{(r)} _{jj'}\cdot \mathbf{1}
\end{array}
\right] \nonumber \\
& = &
\begin{cases}
~\left[
\begin{array}{cc}
-\overline{\mathbf{d}}_{jj'}
& -\mathbf{d}_{jj'} \\
-\overline{\mathbf{s}}^{(-r)}_{jj'}
& -\overline{\mathbf{s}}^{(-r)}_{jj'}
\end{array}
\right] & (s^{(r)}_{jj'} = 1),\\[15pt]
~\left[
\begin{array}{cc}
\mathbf{d}_{jj'}
& \overline{\mathbf{d}} _{jj'} \\
\mathbf{s}^{(-r)}_{jj'}
& \mathbf{s}^{(-r)} _{jj'}
\end{array}
\right] & (s^{(r)}_{jj'} = 0).
\end{cases}
\end{eqnarray*}

From a) and b), consequently, the two absolute columns of
$\widetilde{\mathbf{A}}^{(r)}_{jj'}$ still have the different element(s),
namely, any two absolute column vectors of $\mathbf{A}^{(r)}$ are still not
the same vectors, thus,
$\mu(\mathbf{A}^{(r)}) < 1$.

\section*{Acknowledgment}
This work was partly supported by Japan Ministry of Internal Affairs and
Communications (MIC) in R\&D on Cooperative Technologies and Frequency
Sharing Between Unmanned Aircraft Systems (UAS) Based Wireless Relay
Systems and Terrestrial Networks in 2013.


\end{document}